\pdfoutput=1 
\documentclass[sigconf, balance, nonacm]{acmart}

\usepackage{hyperref}
\usepackage{enumitem}
\usepackage{hologo}
\usepackage{xspace}
\usepackage[ruled,linesnumbered]{algorithm2e}
\SetKw{Continue}{continue}
\usepackage{pifont}
\usepackage{CJKutf8}
\usepackage{makecell}
\usepackage{multirow}
\usepackage{multicol}
\usepackage{tablefootnote}
\usepackage{CJKutf8}
\usepackage{xcolor}
\usepackage[T1]{fontenc}
\usepackage[utf8]{inputenc}
\usepackage{url}
\usepackage{wasysym}
\usepackage{threeparttable}
\usepackage{marvosym}
\usepackage{graphicx}
\usepackage{tcolorbox}
\usepackage{float}
\usepackage{tabularx}
\usepackage[frozencache]{minted}
\usepackage{amsmath}
\usepackage{subcaption}

\newif\ifSimpleMode\SimpleModefalse

\newcounter{ToDo}

\setcopyright{none}
\copyrightyear{2026}
\acmYear{2026}

\acmConference[ICSE 2026]{The 48th International Conference on Software Engineering}{April 12-18, 2026}{Rio de Janeiro, Brazil}
\begin{document}

\newcommand{\SysName}{\textsc{Mars}\xspace}
\title{Towards Scalable and Interpretable Mobile App Risk Analysis via Large Language Models
}

\author{Yu Yang*, Zhenyuan Li*\textsuperscript{\Letter}, Xiandong Ran\textsuperscript{\dag}, Jiahao Liu\textsuperscript{\P}, Jiahui Wang*, Bo Yu\textsuperscript{\ddag}, Shouling Ji*}
\affiliation{%
  \institution{*Zhejiang University, \textsuperscript{\dag}Huawei Technologies Co., Ltd., \textsuperscript{\P}National University of Singapore}, \textsuperscript{\ddag}National University of Defense Technology
  \country{}
}








\renewcommand{\shortauthors}{Trovato et al.}

\begin{abstract}
  Mobile application marketplaces are responsible for vetting apps to identify and mitigate security risks. Current vetting processes are labor-intensive, relying on manual analysis by security professionals aided by semi-automated tools. To address this inefficiency, we propose \SysName, a system that leverages Large Language Models (LLMs) for automated risk identification and profiling. \SysName is designed to concurrently analyze multiple applications across diverse risk categories with minimal human intervention.
To enhance analytical precision and operational efficiency, \SysName leverages a pre-constructed risk identification tree to extract relevant indicators from high-dimensional application features. This initial step filters the data, reducing the input volume for the LLM and mitigating the potential for model hallucination induced by irrelevant features. The extracted indicators are then subjected to LLM analysis for final risk determination. Furthermore, \SysName automatically generates a comprehensive evidence chain for each assessment, documenting the analytical process to provide transparent justification. These chains are designed to facilitate subsequent manual review and to inform enforcement decisions, such as application delisting.
The performance of \SysName was evaluated on a real-world dataset from a partner Android marketplace. The results demonstrate that \SysName attained an F1-score of 0.838 in risk identification and an F1-score of 0.934 in evidence retrieval. To assess its practical applicability, a user study involving 20 expert analysts was conducted, which indicated that \SysName yielded a substantial efficiency gain, ranging from 60\% to 90\%, over conventional manual analysis.

\end{abstract}

\maketitle

\section{Introduction}


Mobile app stores receive thousands of application submissions each day, a significant portion of which pose security risks such as aggressive adware and data-leaking spyware.
These malicious apps steal user privacy and harm the security of the entire mobile ecosystem~\cite{deepraj2024mobile,li2025ui}.
To counter these threats, app store operators invest in heavy application vetting, both before publication and through continuous post-deployment monitoring.
Nevertheless, effective risk identification and profiling remain open challenges.
For example, the prevailing industry practice still relies heavily on manual investigation by security experts --- a process that is slow, costly, and prone to human error~\cite{peck2016analyzing, ogata2018vetting}.
This reliance on human effort stems from two fundamental barriers to automation: the heterogeneity of risks, where diverse threats require distinct detection models, and the heterogeneity of data sources, which complicates the creation of unified analysis pipelines.
As such, existing automated tools~\cite{yang2021pradroid, ibrahim2022method, sutter2024dynamic} are often heavyweight, inflexible, and difficult to maintain, rendering them ineffective at keeping pace with the rapidly evolving threat landscape~\cite{thakur2024android}.


Recent advancements in Large Language Models (LLMs)~\cite{vaswani2017attention, yao2024survey, ge2024openagi} offer a new perspective on mobile app risk profiling.
LLMs excel at understanding and reasoning over complex, heterogeneous data, enabling them to uncover hidden relationships among diverse information sources.
This capability positions LLMs as a powerful tool for comprehensive risk identification and profiling for mobile applications.
However, deploying LLMs to this end is not a trivial task, suffering from several fundamental challenges.
\textit{(a) High-dimensional data:} Comprehensive risk assessment requires synthesizing multiple data sources, including static analysis~\cite{octeau2016combining}, dynamic logs~\cite{graubner2015dynalize}, and metadata~\cite{tashtoush2018analysis}.
However, the computational and context-window constraints of current LLMs make it infeasible to process these features in a single pass, leading to potential information loss and reduced accuracy~\cite{zhang2023siren}.
\textit{(b) Hallucination:} LLMs can generate fluent but factually incorrect outputs, which is particularly problematic in risk assessment, where misjudgments can lead to severe consequences, such as wrongful delisting of applications or potential legal liabilities~\cite{huang2025survey}.



To mitigate these challenges, we propose \SysName, the first LLM-based system for automated and scalable risk identification and profiling of mobile applications.
The core principle of \SysName lies in \textit{pre-organizing high-dimensional data to reduce complexity} and \textit{grounding the LLM's reasoning on this structured knowledge to mitigate hallucination}.
Specifically, \SysName operates on a two-phase pipeline that combines offline knowledge preparation and online real-time analysis to achieve efficient and accurate risk profiling.



\begin{figure}[tbp]
    \begin{center}
        \includegraphics[width=0.48\textwidth]{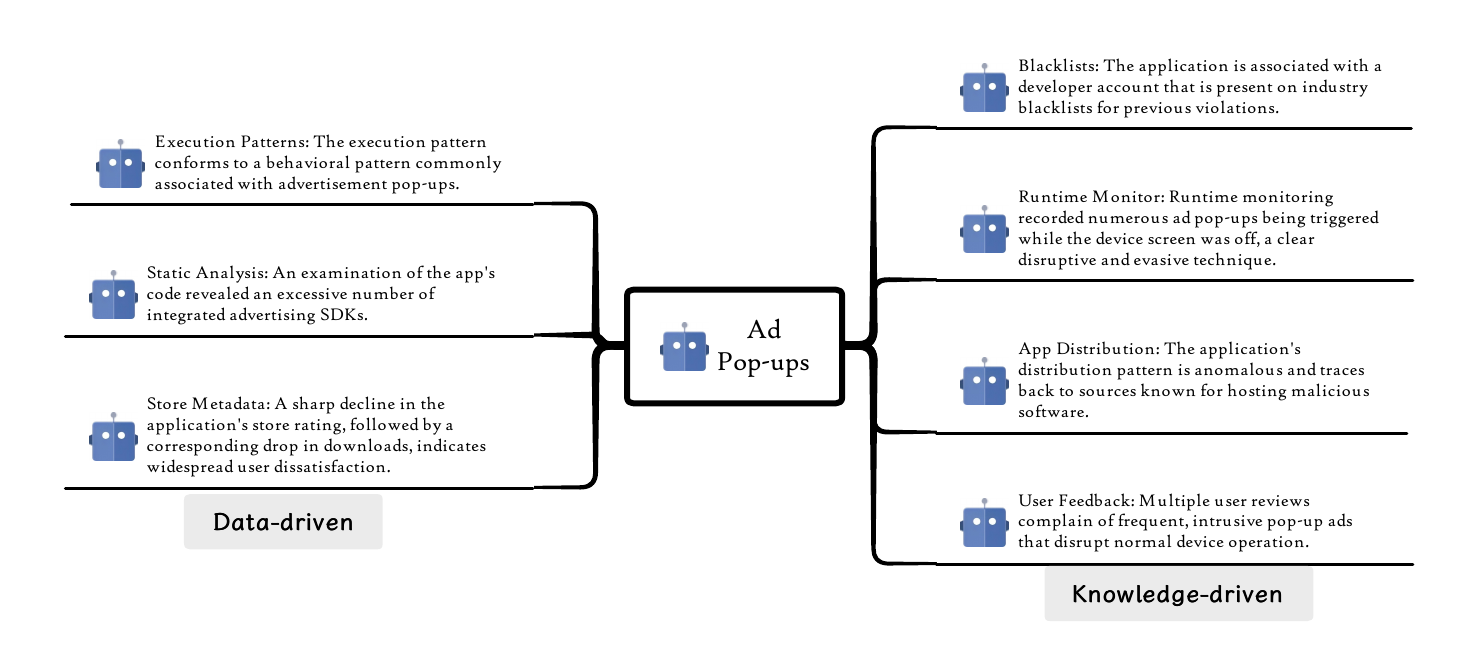}
    \vspace{-0.1in}
    \caption{Risk Identification Tree for Ad Pop-ups}
    \label{fig:decisionTree}        
    \vspace{-0.1in}
    \end{center}
\end{figure}

During the offline phase, \SysName constructs hierarchical knowledge structures --- termed risk identification trees --- to organize risk-relevant features, enabling more effective profiling and reasoning.
For comprehensive risk assessment, \SysName incorporates two complementary sources of information: (1) data-driven signals derived from historical records of delisted mobile apps, based on the observation that apps exhibiting the same type of risk often share similar patterns~\cite{hao2015effective}, and (2) knowledge-driven rules curated by domain experts, whose expertise provides critical insights into subtle risk indicators, such as distribution behaviors~\cite{bobek2019heartdroid}.
Specifically, the risk identification tree is a decision tree structure that captures multiple layers of information, as shown in Figure~\ref{fig:decisionTree}.
The leaf nodes of the tree represent risk-relevant features, while the root nodes correspond to broader risk categories.
It is worth noting that the tree structure plays a crucial role in reducing the complexity of high-dimensional data, enabling \SysName to effectively filter out irrelevant features and focus on the most salient risk indicators.
In the online phase, \SysName leverages the pre-built risk identification tree as a guiding proxy for the LLM's analysis and reasoning, mitigating hallucination and improving the accuracy of risk profiling. 
When a new application is submitted, the system first filters out unrelated risk features based on the tree structure, significantly reducing the input size for the LLM.
The LLM is then prompted to infer the associated risk types and generate a comprehensive report that includes both a machine-readable summary and a human-interpretable evidence chain, as illustrated in Table~\ref{tab:casestudy}.
By structuring the analysis in this manner, \SysName significantly improves efficiency and enables scalable, multi-type risk assessment.



\begin{table}[tbp]
\caption{Risk Report Generated by \SysName} \label{tab:casestudy}
    \vspace{-0.1in}
\footnotesize
\centering
\begin{tabularx}{0.46\textwidth}{c|X}
\toprule
     \textbf{Field} & \textbf{Description} \\ \midrule
     App ID/Name & C11***177 / I Am Music Library \\ 
     Developer & Beijing *** Network Technology Co., Ltd. \\  
     Risk Category & [App Morphing], [Ad Pop-ups] \\ 
     Analysis Time &  2024-12-01 to 2024-12-31 \\ 
     \midrule
    \multirow{10}{*}{\makecell{\textbf{\SysName}\\Evidence\\Snippets}} & 
    \textbf{$\bullet $ User Feedback: }This application is classified as a "Tools" category, but in the user feedbacks, there are risk snippets related to issues such as advertising, game withdrawals, wallet payments, and task withdrawals. \\ 
    & \textbf{$\bullet $ Execution Patterns: }This app matched the morphing mode, with the number of launches changing from 2469 to 3084. The list of unrelated packages launched by this app is as follows: \texttt{com.asdg.xwdd}, \texttt{com.aaqe.jymtf}, ... \\
    & \textbf{$\bullet $ Runtime Monitor: }The endpoint information shows that during the application's operation, the average number of pop-ups was 20.95 times, and the average screen-off usage time was 20,109 seconds. \\
    & \textbf{$\bullet $ App Distribution: }The distribution of this application includes C10***531, C10***401, C11***363 and C11***535, with C11***363 being a removed game application. \\
    & \textbf{$\bullet $ Anti-virus Engine: }The engine finds multiple Ad SDK implants, including \texttt{com.by***ance.pa**le.sdk}, \texttt{com.b**du.mo**ds.sdk}, \texttt{com.k**d.sdk}, ... \\ 
    \midrule
     \multirow{6}{*}{\makecell{\textbf{\SysName}\\Risk Description}} & As a tool-based application, there are user comments mentioning games, withdrawals, and ad pop-ups. Additionally, the caller-callee analysis matched the App morphing pattern, as the application triggered many other Java packages. Furthermore, this app distributed a previously removed game application. Based on these factors, we can conclude that the app exhibits a [App Morphing] risk. Moreover, endpoint monitoring data reveals multiple pop-up occurrences, and the Malicious Engine has reported the activation of several ad-related SDKs. Therefore, there are reasonable grounds to believe that this App also poses a [Foreground Ad Pop-ups] risk. \\  
     \midrule
     \multirow{3}{*}{\makecell{Risky App with\\ similar pattern}} & 
     1) C11***339/Move a brick. Risk Category: [App Morphing]. 2) C11***351/Thousand Miles of Journey. Risk Category: [Ad Pop-ups].     \\ \bottomrule
     
\end{tabularx}
\vspace{-0.1in}
\end{table}

To systematically evaluate the performance and practical utility of \SysName in identifying and reasoning about mobile application risks, we conducted a comprehensive study on a dataset including 2,232 real-world mobile apps.
Quantitative analysis on a ground-truth labeled subset showed high accuracy, with a risk identification F1-score of 0.838 and an evidence retrieval F1-score of 0.934.
The system is also highly efficient, with an average analysis time of 24.68 seconds and an estimated cost of just \$0.015 per application.
Furthermore, a user study involving 20 security practitioners confirmed \SysName's practical value. Participants experienced a 60\% --- 90\% improvement in analysis efficiency, and the system's generated evidence chains received high ratings for correctness, integrity, and clarity.
These results underscore \SysName's real-world applicability and interpretability.

In summary, this paper makes the following contributions:

\begin{list}{\labelitemi}{\leftmargin=1.6em}
 \setlength{\topmargin}{0pt}
 \setlength{\itemsep}{0em}
 \setlength{\parskip}{0pt}
 \setlength{\parsep}{0pt}
    \item \textbf{The first LLM-driven system for scalable Android application risk profiling.} To the best of our knowledge, \SysName is the first system to automate the analysis of multiple risk types from diverse data sources in a unified, LLM-based framework, significantly enhancing the accuracy and efficiency of the process.
    
    \item \textbf{A novel knowledge-guided framework for LLM-based analysis.} We propose a hybrid framework that uses a hierarchical knowledge tree, built from expert rules and historical data, to ground the LLM's reasoning. This structure mitigates common LLM challenges, including hallucination and difficulty with high-dimensional inputs, enabling accurate and context-aware risk assessment.

    \item \textbf{A comprehensive evaluation on real-world data.}
    We demonstrate the practical viability of \SysName through extensive experiments on a large-scale dataset. The results confirm its high accuracy for risk identification and evidence generation, as well as its operational efficiency in terms of both time and cost.
    To facilitate future research, we plan to open-source our work at \url{https://anonymous.4open.science/r/MARS-FEC4}.

    
\vspace{-\topsep}
\end{list}


\section{Background and Related Work}
\label{sec:background}




\subsection{Application Risks Analysis}


The rapid proliferation of mobile devices has driven the development of millions of applications.
While these apps provide a wide range of functionalities, they also introduce numerous critical security threats~\cite{faruki2014android, ahmed2017android,liu2024unraveling}. 
Table~\ref{tab:majorRisk} summarizes the primary categories of mobile application risks that have garnered significant attention and form the focus of this study.
To better illustrate these risks, we also present their top three associated feature groups; further details are provided in \S\ref{sec:sec4}.
Over the past decade, researchers have developed various detection systems to counter these threats~\cite{huang2020detecting, mayrhofer2021android, gopinath2023comprehensive}.
Most of these systems are specialized tools designed to detect a specific type of risk, such as permission misuse, malware, or ad fraud.
For instance, tools like Whyper~\cite{pandita2013whyper} and FlowDroid~\cite{arzt2014flowdroid} examine permission usage. DroidChameleon~\cite{rastogi2013droidchameleon} and its successors~\cite{alam2016droidclone, garcia2018lightweight} use code similarity to detect copied applications.
Others, like MAdFraud~\cite{crussell2014madfraud}, analyze user-interaction data to identify ad fraud.

\begin{table*}[htbp]
\footnotesize
    \centering
    \caption{Types and Description of Major Application Risk of Concern}  
        \begin{tabularx}{0.98\textwidth}{l|X|l}
        \toprule 
             Risks Name & \multicolumn{1}{c|}{Description} & Top-3 Related Feature Groups \\
        \midrule 
            Ad Pop-ups & Advertisements that appear prominently while the app is in active use, disrupting the user experience and core functionality. & User Feedback, Execution Patterns, Dynamic Load \\
            Unexcepted Pop-ups & Pop-ups that surface without user initiation, often a sign of adware, phishing attempts, etc. & Runtime Monitor, Static Analysis, Execution Patterns \\
            Retention & Behaviors allowing apps to covertly re-enable themselves or persist after being uninstalled. & Runtime Monitor, Dynamic load, Store Metadata \\
            App Morphing & App alters its functionality post-installation to reveal malicious features that were hidden during the review process. &  User Feedback, Execution Patterns, Discrepancy\\
            Illegal Features & Functionalities that violate platform policies or laws (e.g., unauthorized data collection). & User Feedback,  Store Metadata, Blacklist\\
            Content Risk & Presence of harmful, inappropriate, or misleading material that may jeopardize user safety, privacy, or regulatory compliance. & User Feedback, Screenshot, Store Metadata\\
            App Counterfeiting & Counterfeit Apps to mislead users, potentially distributing malware or infringing intellectual property. & App Similarity, User Feedback, Store Metadata\\
            Malware & Apps that are designed to harm devices, steal sensitive data, or exploit vulnerabilities. & Runtime Monitor, App Distribution, Network Feature \\
            
        \bottomrule
        \end{tabularx}
    \label{tab:majorRisk}
\end{table*}


However, this specialization has become a major obstacle to large-scale application risk profiling.
In particular, the reliance on narrowly focused, single-purpose tools presents two fundamental challenges for platform operators.
First, maintaining a diverse set of tools --- each with its own data dependencies and operational requirements --- is fragile and difficult to sustain.
Second, such systems often lack interpretability, making it challenging for security analysts to trace and validate their decisions.
As such, building a comprehensive defense by integrating these disparate tools introduces significant operational complexity and leads to unsustainably high long-term costs.

\begin{figure}[tbp]
    \centering
    \includegraphics[width=0.42\textwidth]{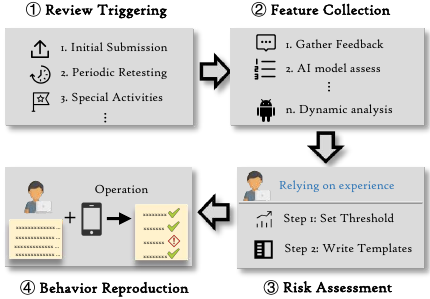}
    \caption{Manual App Risk Review Process}\label{fig:reviewProcess}
    \vspace{-0.1in}
\end{figure}

Currently, mobile app platforms still rely heavily on manual review, where security engineers use their expertise to assess risk.
A typical review follows four stages, as Figure~\ref{fig:reviewProcess} shows, namely, review triggering, feature collection, risk assessment, and behavior reproduction.
Reviews may be initiated due to new app submissions, periodic audits, runtime anomalies, or user feedback.
Once triggered, the platform utilizes a collection of static and dynamic analysis tools to extract behavioral and structural features from the application under review.
These tools collectively extract hundreds of high-dimensional features from each application.
Security engineers are then responsible for defining threshold conditions to identify anomalous features, drafting corresponding risk report templates, and reproducing the identified risks on mobile devices.
These steps collectively form a complete app profiling workflow, encompassing detection, documentation, and verification.

As for existing automated models, they are hampered by a fragmented design. 
Individual detectors operate in silos, each analyzing a narrow set of features, which prevents them from forming a complete picture of an app's behavior or handling mixed data sources effectively.
Accordingly, platforms must maintain many separate, manually-tuned models for each risk category, leading to significant engineering costs and is fundamentally difficult to scale~\cite{sutter2024dynamic}.

\subsection{LLMs in Security Analysis}

Recent advances in large language models (LLMs) have demonstrated remarkable capabilities in language understanding~\cite{kim2024understanding}, logical reasoning~\cite{wang2023can}, and generalization~\cite{wu2023next},  offering powerful new tools for the field of cybersecurity~\cite{xu2024large}.
A number of security-oriented systems have begun to explore this potential: Crimson~\cite{jin2024crimson} leverages LLMs for threat detection and attribution, RCACopilot~\cite{chen2024automatic} applies them to cloud incident diagnosis, AdbGPT~\cite{feng2024prompting} enables Android bug reproduction via prompt interaction, and GPTScan~\cite{sun2024gptscan} uses LLMs to uncover source-code vulnerabilities. 


Recent studies have demonstrated the potential of large language models (LLMs) in tasks such as threat detection and source code analysis.
Owing to their capacity to process large volumes of heterogeneous data, LLMs are well-positioned to advance mobile application risk profiling.
However, their direct application to this domain faces two significant challenges: the inherent limitation in input context length and the tendency to generate hallucinated content.
To overcome these obstacles, we propose a knowledge-based risk identification tree.
This tree serves as a critical pre-processing mechanism that distills high-dimensional features into concise, risk-relevant signals.
By doing so, it simultaneously addresses the context length constraint—through input reduction—and mitigates hallucination by grounding the LLM's analysis in structured, domain-specific knowledge.

\subsection{Motivating Example}

Existing manual review workflows are fragile, often overwhelmed by alerts and reliant on single evidence sources with poor cross-validation. The case of the ``Pocket Accounting'' app in Table~\ref{tab:casestudy} is a clear example. Initially flagged based on vague user feedback about ads, the review required a time-consuming manual effort to simply reproduce the behavior. The eventual delisting decision, dependent on this single piece of evidence, was slow and lacked comprehensive validation.

\SysName replaces this fragile process with a robust, automated framework. For the same Pocket Accounting app, \SysName automatically synthesized five distinct evidence sources: user feedback, code execution patterns indicating app morphing, idle-hour activity logs, distribution links to previously delisted apps, and alerts from commercial antivirus engines. It then cross-validated these findings to produce a single, interpretable evidence chain that correctly identified two separate risks: "App Morphing" and "Foreground Ad Pop-ups." By replacing arduous manual reproduction with automated, multi-faceted evidence synthesis, \SysName delivers a faster, more precise, and transparent risk assessment.

\section{System Overview}

\begin{figure}[tbp]
    \centering
    \includegraphics[width=0.48\textwidth]{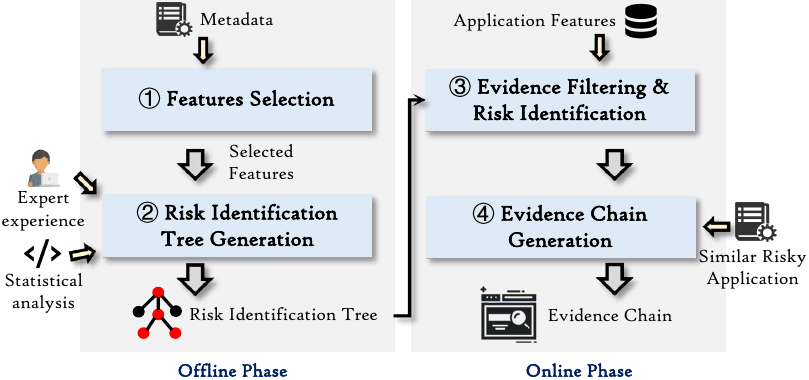}
    \caption{System Overview}\label{fig:systemOverview}
\end{figure}

The overall workflow of \SysName is provided in Figure~\ref{fig:systemOverview}.
It consists of two main phases to embed expert knowledge and statistical patterns into a real-time analysis pipeline. 

In the offline phase, \SysName constructs hierarchical knowledge trees to guide its online risk analysis. To enhance the robustness and domain relevance of the tree, the construction process is guided by two complementary strategies: a knowledge-driven approach using heuristics from expert security engineers, and a data-driven approach using statistical signals from a curated set of delisted applications. These dual foundations enable the system to capture both explicit and latent risk indicators observed in real-world mobile threats. Furthermore, the hierarchical knowledge trees serve as a contextual grounding layer for the LLM, significantly reducing hallucination and improving interpretability.
S
The first step in the offline phase is raw feature extraction and dimensionality filtering, because the raw features from these sources are high-dimensional and often noisy, the process begins with an LLM-driven feature filtering module that eliminates irrelevant data to reduce computational load and improve model focus~\cite{white2023prompt, schulhoff2024prompt}. Following this filtering stage, we aggregate features that share similar semantics and exhibit consistent decision patterns into unified feature groups. Then, we perform statistical analysis on historical risk cases to construct a bipartite graph that links each feature group to one or more specific risk categories. \SysName builds the tree by creating leaf nodes, where each node is a specialized, LLM-based agent responsible for analyzing a specific feature group. The final tree enables robust and interpretable risk classification by integrating and correlating the findings from these distributed agents.

Leveraging the risk identification tree generated in the offline phase, \SysName conducts real-time risk profiling for real-world applications. The online pipeline begins with data collection and preprocessing, where multi-source data is gathered, cleaned, and categorized into unstructured features (e.g., descriptions, logs) and structured features (e.g., API call frequencies).
Next, these features enter an evidence filtering module. Here, an LLM-agent analyzes unstructured data for semantic risk patterns, while statistical models find anomalies in the structured data. The resulting signals are then passed to the risk identification module, which uses the pre-constructed knowledge tree and a historical database to assign precise risk labels.
Finally, an evidence chain generation module synthesizes all validated findings, including risk snippets, metadata, and precedents, into a coherent report. As shown in Table~\ref{tab:casestudy}, these reports clearly articulate the rationale behind each decision, ensuring the entire risk assessment process is systematic, transparent, and directly actionable for app store moderation teams. \SysName adopts a lightweight design and incurs minimal overhead, making it well-suited for large-scale, high-volume deployment scenarios such as mainstream app stores.





\begin{table}[tbp]
\Small
    \centering
    \caption{Feature Groups (Not Exhausted)} \label{tab:featureGroups}
    \begin{tabularx}{0.48\textwidth}{c|X}
        \toprule
         & Feature Groups   \\ 
        \midrule
        \makecell{Knowledge-driven} & Blacklist, Runtime Monitor, App Distribution, User Feedback, Discrepancy, Screenshot, App Similarity, Network Feature, Anti-virus Engine\\
        \makecell{Data-driven} & Execution Patterns, Usage patterns, Static Analysis, Dynamic Load, Store Metadata, Update frequency \\
        \bottomrule
        \end{tabularx}
    \vspace{-0.1in}
\end{table}

\section{Risk Identification Trees Construction}
\label{sec:sec4}

To enable interpretable and structured risk profiling, we designed a hierarchical risk identification tree composed of three node types, each with a specialized agent.
The root node represents a high-level risk category, as listed in Table~\ref{tab:majorRisk}; its agent aggregates findings from lower-level nodes to make a final classification decision.
Intermediate nodes correspond to semantic feature clusters (e.g., runtime behavior), and their agents are responsible for analyzing that specific class of data for anomalies.
Table~\ref{tab:featureGroups} details the feature groups used for our risk analysis. These features were selected for their high relevance to the risks under investigation, and our system is designed to be extensible, allowing more feature groups to be added at a minimal cost.
Leaf nodes represent the raw feature dimensions (e.g., number of background wakeups) that serve as the initial inputs to the tree.

This tree is built in a multi-step offline process. First, raw features are classified as either knowledge-driven (from expert heuristics) or data-driven (from historical data). After low-utility features are pruned, the remaining dimensions are grouped into the semantic clusters that define the intermediate nodes. Finally, these clusters are mapped to the appropriate risk categories (the root nodes) using a bipartite graph, which forms the tree's final structure.

\subsection{Knowledge-Driven Features}
To ground our system's analysis in practical expertise, we developed a set of knowledge-driven features that encode the heuristics used by experienced security analysts. These features are grouped into several key categories, each designed to identify deviations from normal application behavior, for example:
(1) User Feedback. To process direct user reports from sources like app store reviews, we designed a specialized agent. Using prompt engineering with role-playing and few-shot examples, this agent filters raw feedback, extracts risk-relevant snippets, and classifies them into predefined categories (e.g., advertising issues, content anomalies).
(2) Runtime monitor. This category includes data captured from live monitoring, such as ad pop-ups while the screen is off or unusual background activity, that reveals covert or malicious patterns often missed by static analysis.
(3) Discrepancy. This focuses on mismatches between an application's declared purpose and its actual functionality. Such inconsistencies are common indicators of deceptive practices like app morphing.
(4) App Similarity. This measures an application's resemblance to previously delisted or blacklisted apps in terms of UI design, code structure, or icons, enabling the effective detection of counterfeits.

To develop and validate these feature categories, we conducted workflow ethnography with on-call security engineers, combining their insights with the statistical analysis of historical risk reports and delisted applications. This rigorous process formalizes real-world operational expertise into the structured knowledge base that forms the semantic foundation of our risk identification tree.

Furthermore, \SysName assigns a dedicated LLM-based agent to each feature group to analyze it for anomalies. The agent for User Feedback, for instance, is crucial because users often expose nuanced issues that automated static or dynamic analysis can miss. As Prompt \uppercase\expandafter{\romannumeral1} shows, to process this valuable data, we designed a specialized agent that leverages advanced prompt engineering with role-playing and few-shot learning strategies. This agent systematically processes raw user comments to first filter out irrelevant noise, then extract risk-relevant evidence snippets, and finally classify them into predefined risk factors (e.g., advertising, content anomalies, functional issues). This targeted, agent-based approach enables the efficient extraction of actionable insights from large volumes of unstructured user feedback, enhancing the overall reliability and scalability of our risk profiling system.

\begin{figure}[tbp]
\begin{tcolorbox}[colback=gray!10, colframe=black, title=Prompt \uppercase\expandafter{\romannumeral1}: User Feedback Analysis]
\label{prompt:p1}
\Small

As an expert user feedback reviewer, your task is to identify and extract potential risk snippets from the user feedback provided below. You must then assign the single most relevant risk factor to the feedback, choosing from the predefined list.
There are some risk factors you should know: 

\vspace{0.1in}
\textbf{Predefined Risk Factors:}
\begin{minted}[fontsize=\footnotesize]{text}
  1. Advertising-related: ...
  2. Content Anomaly: ...
  ...
\end{minted}

\textbf{Example of Input and Required Output Format:}
\begin{minted}[fontsize=\footnotesize]{json}
Input: {
  1. "It's terrible, when I opened it, the application con-
  stantly pops up ads."
  2. "It was fine before, but now it's all ads, and it aff-
  ects my experience."}
Output: {
  "Quality": "High",
  "Tendency": "Negative",
  "RiskInfo": {
    "Snippets": 
    "The application constantly pops up advertisements.
    | There are advertisements, which affect the experience.
    |...",
    "Risk Factor": "Advertising-related" }}
\end{minted}

\textbf{Data to Analyze:} \underline{<Input Data>}

\vspace{0.1in}
\textbf{LLM Output:} \underline{<Output Result>}

\end{tcolorbox}
\vspace{-0.1in}
\end{figure}

\subsection{Data-Driven Features}


To complement our knowledge-driven features, we employ data-driven methods to identify statistical patterns indicative of risk. This involves analyzing time-series and cross-sectional data from large historical case repositories to detect significant deviations from normal behavior.

\begin{figure}[tbp]
    \centering
    \includegraphics[width=0.42\textwidth]{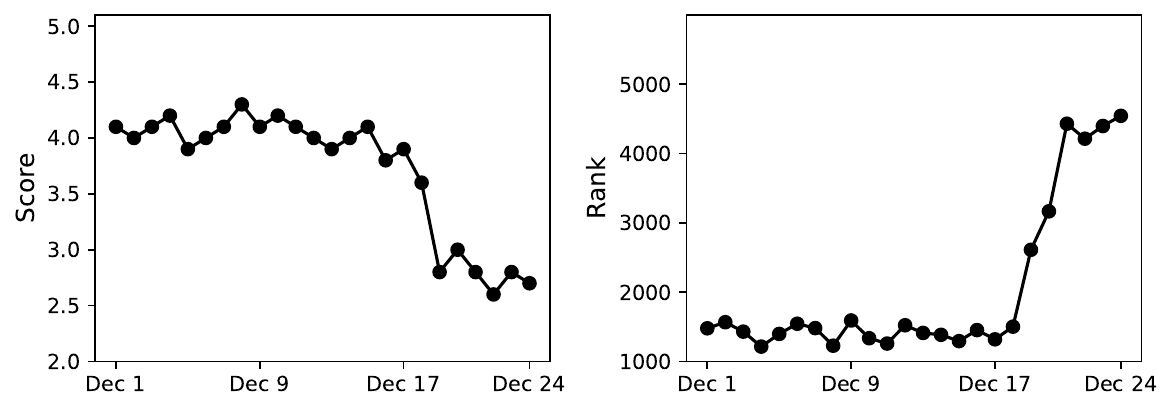}
    \vspace{-0.05in}
    \caption{Temporal fluctuations in app rating and ranking}\label{fig:data_driven}
    \vspace{-0.1in}
\end{figure}

The first strategy is an intra-application analysis, where we examine temporal fluctuations in an app's features. As shown in Figure~\ref{fig:data_driven}, this allows us to capture events like sudden drops in user ratings that may signal an emerging risk. We ensure the validity of this analysis by carefully selecting time intervals based on statistical distributions to balance data freshness and consistency.

The second strategy is an inter-application comparison between risky and benign apps. This helps identify feature dimensions, such as abnormal API call frequencies, that consistently diverge between the two groups. Together, these comparative analyses support a final round of feature dimensionality reduction, enabling us to isolate a compact set of the most salient statistical indicators for risk profiling.

In our data-driven feature selection process, several features require threshold-based evaluation. We adopt two thresholding methods depending on the nature of the feature:
For dynamic features with relatively stable distributions (e.g., load frequency, update intervals), we establish thresholds using descriptive statistics like the mean and median. These are computed for specific application categories to account for normal variations between different types of apps.
For metrics with high variance, such as download rankings or user scores, we identify outliers using a 95\% confidence interval. For example, to detect a significant fluctuation in an app's ranking, we calculate its expected ranking interval based on the previous week's data using Equation (1). A current rank falling outside this interval is flagged as a statistical anomaly, serving as a potential early warning signal.
\begin{equation}
    \small CI = \bar{x} \pm t_{\alpha/2} \cdot \frac{s}{\sqrt{n}},
\end{equation}  
where $CI$ represents the confidence interval, $t_{\alpha/2}$ is the critical value for a given confidence level, $s$ is the standard deviation, and $n$ is the sample size.

This analysis pipeline allowed us to identify over a dozen statistically significant features, including: 
(1) Execution Patterns: Runtime behaviors, including API call frequencies, caller-callee relationships, and interaction patterns.
(2) Dynamic Load: The runtime inclusion of external components, measured by the frequency and source of dynamically loaded packages.
(3) Blacklist: Indicators linking the developer, application, or download source to known industry blacklists.
(4) Store Metadata: Market-level attributes such as download ranks, user ratings, and the presence of a privacy policy, which can signal reputation or abnormal activity.
These data-driven features, validated through intra- and inter-application comparisons, enable the scalable identification of high-impact risk factors that complement our knowledge-driven heuristics.

\section{Risk Identifying and Reasoning }

The risk identification tree enables a systematic, three-stage process for profiling application risks. The process begins with evidence filtering, where the pre-built tree processes raw application features to extract relevant snippets. 
This is followed by risk identification, where a dual-mechanism approach confirms the risk category based on the filtered evidence. 
Finally, the report generation stage synthesizes these findings into comprehensive risk reports, complete with evidence chains and similar risky app confirmed, to facilitate final confirmation by security analysts.


\subsection{Evidence Filtering}

Our process for filtering raw features into meaningful evidence consists of two main stages. First, we organize all features into predefined groups, as summarized in Table~\ref{tab:featureGroups}. Second, we employ a system of specialized, LLM-based agents, with a dedicated agent assigned to each group for analysis. We adopt this approach because rule-based strategies are inadequate for handling the complexity of textual data like user feedback (handled by the User Feedback Analysis Agent) and application profiles. Each agent, therefore, leverages large language models to perform its tailored discriminative function.

This agent-based design provides three key advantages. First, the agents selectively analyze relevant feature groups, preventing the core model from being overwhelmed by massive, high-dimensional inputs. Second, they distill raw data into semantically meaningful summaries, reducing the amount of data for subsequent analysis. Third, targeted screening reduces the illusion of large models, allowing for more accurate subsequent risk identification and report generation. 

For each feature group, we design a tailored prompt that reflects the unique semantics and risk signals associated with that group. In the critical reasoning chain construction, we simulate the decision-making process of expert analysts, enabling the model to align its inference flow with human domain expertise. Furthermore, we iteratively refine and adapt the prompts based on the model’s intermediate reasoning outputs, ensuring both interpretability and task-specific performance. For instance, the Discrepancy Agent uses a custom-engineered prompt to detect mismatches between an app's declared category and its description, a common indicator of app morphing. This targeted analysis on a focused data slice exemplifies how our multi-agent system achieves precise and contextually relevant judgments.

\subsection{Risk Identification}
\label{subsec:validation}

Identifying a risk based solely on a single agent and limited evidence snippets will result in a high false alarm rate. 
To address this, we design two verification mechanisms to validate the results of risk identification, providing the identification tree with a more comprehensive perspective and improving its accuracy.

The first validation mechanism is integrated directly into the identification tree. A risk identified by the identification tree is forwarded to subsequent stages only if at least two mutually verifiable leaf nodes are triggered.
Based on an analysis of agent hit distributions across a large dataset of delisted applications, we design dozens of agent validation rules. For example,  the ``Runtime Monitor'' contains snippets related to screen-off usage, which is often corroborated by corresponding anomalous usage periods in the ``Usage patterns'' for validation. Besides, ``User Feedback'', such as comments mentioning pop-ups, often reveals that, through ``Exection patterns'' the application initiates other packages unrelated to its core functionality.

\begin{figure}[tbp]
\begin{tcolorbox}[colback=gray!10, colframe=black, title=Prompt \uppercase\expandafter{\romannumeral2}: Retrieve Similar Pattern]
\label{prompt:p2}
\Small

As an expert application reviewer, your task is to determine if the target application is similar to any of the provided historical records, based on the following criteria.

\vspace{0.1in}
\textbf{Definition of "Similar":}
\begin{minted}[fontsize=\footnotesize]{text}
  An application is considered similar if it shares at least 
  two identical risk indicators with the target application.
\end{minted}

\textbf{Historical Records from Delisted Database (via RAG):}
\begin{minted}[fontsize=\footnotesize]{json}
  "C17**21": ["anomalous-rank", "comment", "malicious-callee"],
  "C12**65": ["anomalous-rate", "ad pop-ups"],
  "C18**23": ["malicious-activate", "ads-system-bar"]
\end{minted}

\textbf{Target Application for Analysis:}
\begin{minted}[fontsize=\footnotesize]{json}
  "C16**37": ["anomalous-rank", "ad pop-ups", "malicious-call-
  callee"]
\end{minted}

\vspace{0.1in}
\textbf{LLM Output}: \underline{"C17**21": ["anomalous-rank", "comment", "malicious-callee"]}

\end{tcolorbox}
\vspace{-0.1in}
\end{figure}


Subsequently, recognizing that applications of the same risk category often exhibit similar anomaly patterns, we introduced a second verification mechanism. We formulated the prompt schema as illustrated in Prompt \uppercase\expandafter{\romannumeral2}, after the identification tree provides the risk category and the evidence snippets, we search our delisted database for records with a similar distribution of evidence snippets. This process confirms the consistency of the identified risk with historical data, enhancing the reliability of our identification.

To implement this design, we utilize Retrieval-Augmented Generation (RAG) process to cross-reference a database of previously delisted applications. Specifically, we employ the \textit{bge-large-zh-v1.5} embedding model~\cite{bge_embedding} to calculate embeddings for the evidence snippets of the applications under evaluation. We then use cosine distance to retrieve the top three closest records from the corresponding risk database. Subsequent verification is conducted by the large language model, as shown in Prompt \uppercase\expandafter{\romannumeral2}.
This process not only ensures that the detected risks are in line with known patterns but also leverages advanced NLP techniques to improve the accuracy and efficiency of our risk verification methodology.

\subsection{Evidence Chain Generation}
\label{subsection:risk_report_generation}

When an application is confirmed to pose security risks, it is forwarded to the evidence chain generation module, which is designed to produce alert outputs that are both informative and easily interpretable by human reviewers. We incorporate template examples into the prompts to guide the model’s learning process. By observing and internalizing these structured demonstrations, the LLM exhibits enhanced capability in describing logical relationships between risk-related features, leading to more coherent and interpretable reasoning chains during risk assessment. To accommodate different usage contexts and stakeholder needs, we provide two formats for alarm output: a structured JSON file and a human-interpretable narrative report.

For structured output, we generate a standardized JSON file using a carefully designed prompt.
\ding{172} Application Overview. Includes essential metadata such as the application name, ID, and other identifiers;
\ding{173} Risk Category Summary. States the risk type(s) associated with the application;
\ding{174} Evidence Chain Details. Contains a comprehensive list of evidence snippets, including raw feature data and the corresponding analytical logic used to derive risk conclusions;
\ding{175} Similar Delisted Applications. Provides references to historically removed apps that exhibited similar risk patterns, offering contextual grounding and aiding in comparative analysis.

In addition to the structured format, we also generate a human-interpretable narrative report using a dedicated prompt 
built for LLMs. This prompt leverages few-shot learning with multiple curated examples to transform discrete evidence snippets into a cohesive and fluent textual summary. This natural language summary improves interpretability and enables security engineers to quickly grasp the rationale behind the system’s risk assessments. By offering these dual formats, \SysName ensures both machine-readability for downstream automation and intuitive comprehension for manual review processes.

\section{Evaluation}
In this section, we focus on evaluating the performance of \SysName by answering the following research questions (RQs)

\vspace{-\topsep}
\begin{enumerate}[label=\textbf{RQ\arabic*.}, left=0pt, itemsep=0em]
\item How accurate is \SysName in risks identification and evidence retrieving? (\S\ref{eval:accuracy})
\item What is the computational cost of \SysName? (\S\ref{eval:efficiency})
\item Can \SysName enhance the efficiency of manual analysis? (\S\ref{eval:applicability})
    
\end{enumerate}
\vspace{-\topsep}

\subsection{Evaluation Setup}




The experiments were conducted on a dedicated computational node equipped with an 8-core CPU, 64GB of RAM, and an NVIDIA V100 GPU to accelerate LLM inference. The \SysName is implemented with approximately 2,500 lines of Python code.

\subsubsection{Ethical Considerations.} All experiments were conducted within the company's secure internal network, utilizing LLMs deployed on-premise. All data was processed locally, and no information was transmitted to any external parties. This approach was implemented to fully safeguard the rights of both developers and users.

\subsubsection{LLM Selection.} The selection of LLMs was guided by practical deployment constraints, including data privacy, inference speed, and computational overhead. These requirements prohibit the use of proprietary, cloud-based models like GPT-4, as sensitive application metadata cannot be exposed to third-party services. Consequently, our study focuses on locally-deployable, open-source models. The models evaluated include compact variants of {Llama3}, {Qwen2.5}, and {DeepSeek-R1}. 
For our main evaluation, we utilized a fixed LLM temperature of 0.7. This parameter was empirically validated to provide the best trade-off between creative diversity, necessary for identifying new risk patterns, and the deterministic consistency essential for reproducible results.

\subsubsection{Datasets.} 
\begin{figure}[!t]
    \centering
    \begin{subfigure}[b]{0.5\linewidth}
        \centering
        \includegraphics[width=\linewidth]{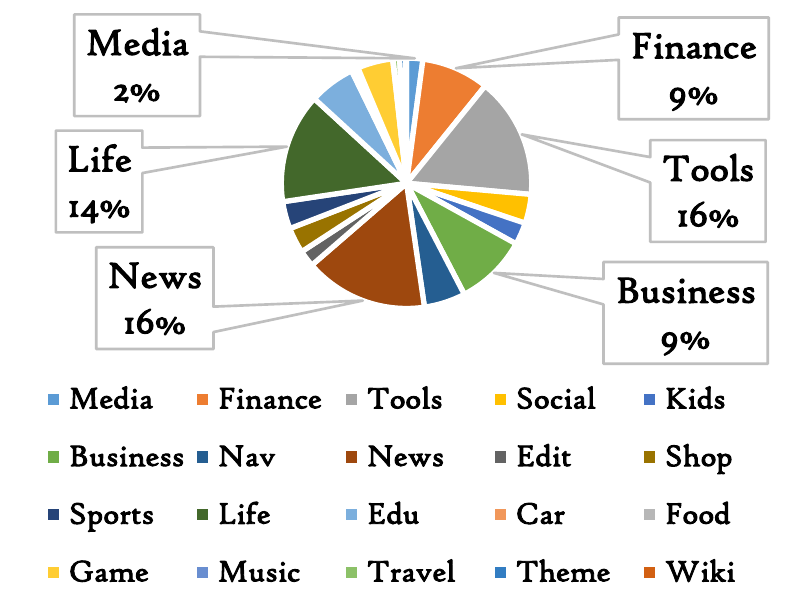}
        \caption{App Category Statistics}
        \label{fig:d1}
    \end{subfigure}%
    \hfill
    \begin{subfigure}[b]{0.49\linewidth}
        \centering
        \includegraphics[width=\linewidth]{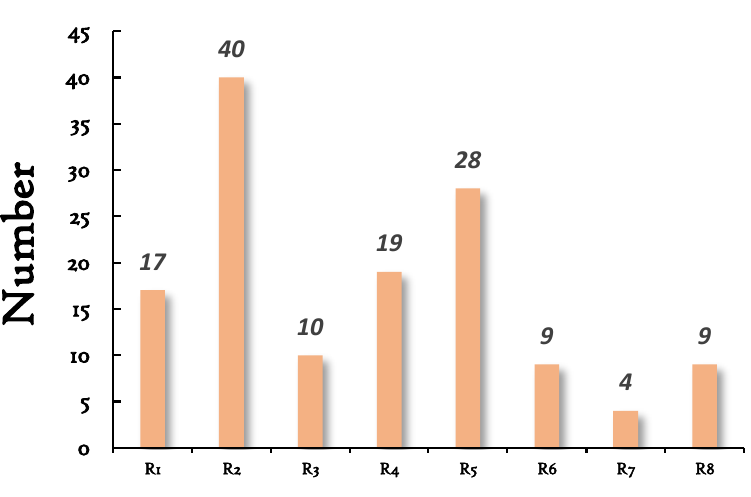}
        \caption{Risk Type Statistics}
        \label{fig:d2}
    \end{subfigure}
    
    \caption{Statistics of the Evaluation Dataset}
    \vspace{-0.15in}
\end{figure}

Our evaluation utilizes a dataset of 2,232 applications collected from a mainstream Android App stores. This dataset was partitioned into two distinct subsets to facilitate a comprehensive evaluation:
1) Labeled Risk Dataset: A subset of 136 applications was manually annotated with ground-truth risk labels. This dataset is used to quantitatively measure the system's effectiveness, particularly for calculating metrics like the False Negative Rate (FNR). The distribution of risk categories within this set is detailed in Figure~\ref{fig:d2}.
2) Unlabeled Real-World Dataset: The remaining 2,096 applications were left unlabeled to simulate a real-world analysis scenario. This set is used to assess the system's performance on unseen data and to calculate the False Positive Rate (FPR) through post-hoc manual analysis of flagged applications. The categorical distribution of these apps is shown in Figure~\ref{fig:d1}.

\subsubsection{Baseline Approaches.} 
Existing state-of-the-art (SOTA) methods for app security focus on the in-depth analysis of single-risk detection with a single data source, such as in privilege escalation detection based on dynamic analysis, and rely on manual analysis for subsequent confirmation and remediation~\cite{tebib2023privdroid}. 
In contrast, \SysName leverages Large Language Models (LLMs) to fuse diverse data sources for simultaneous multi-risk analysis. It automatically summarizes the reasoning process behind its judgments, a capability not present in prior work. As an effective complement to previous approaches, our system significantly enhances the efficiency of manual analysis.

Given these fundamental differences in scope, multi-risk vs. single-risk, data fusion vs. single-source, and automated reasoning vs. manual confirmation, a direct, like-for-like comparison with SOTA methods is not meaningful. We therefore adopt the foundational LLMs as our primary baseline. The contributions of our unique architecture are then validated through extensive ablation studies, which systematically evaluate our domain-specific data processing pipeline and multi-risk analysis framework.

\definecolor{fncolor}{rgb}{0,0,0}
\definecolor{fpcolor}{rgb}{0,0,0}
\begin{table*}[t]
\caption{Accuracy of Risk Identification and Evidence Selection ($\textcolor{fncolor}{-false\_negative} /\textcolor{fpcolor}{+false\_positive}$)} \label{tab:eva1}
\scriptsize
\centering
    \begin{tabular}{@{}l|cccccccc|cccc@{}}
    \toprule
         & \multicolumn{8}{c|}{\# of Risk/Evidence Identified} & \multicolumn{3}{c}{Overall} \\
         \textbf{}       & R1 & R2 & R3 & R4 & R5 & R6 & R7 & R8  & Precision & Recall & F1-Score\\ 
    \midrule
         \multicolumn{12}{c}{Risk Identification Accuracy} \\ \midrule 
    \# of Risks (GT)   &  17 & 40   &  10  &  19  &  28  &  9  &  4    &  9 & 1.000 & 1.000 & 1.000   \\ \midrule
    Llama3-8B    & \textcolor{fncolor}{-0} (\textcolor{fpcolor}{+17})  &  \textcolor{fncolor}{-0} (\textcolor{fpcolor}{+38})  &  \textcolor{fncolor}{-0} (\textcolor{fpcolor}{+10})  &  \textcolor{fncolor}{-0} (\textcolor{fpcolor}{+19})  &  \textcolor{fncolor}{-0} (\textcolor{fpcolor}{+28})  &  \textcolor{fncolor}{-0} (\textcolor{fpcolor}{+8})  &   \textcolor{fncolor}{-0} (\textcolor{fpcolor}{+4})   & \textcolor{fncolor}{-0} (\textcolor{fpcolor}{+3})  & 0.066 & \textbf{1.000} & 0.124   \\
    Qwen2.5-7B     & \textcolor{fncolor}{-0} (\textcolor{fpcolor}{+15})  & \textcolor{fncolor}{-0} (\textcolor{fpcolor}{+36})   & \textcolor{fncolor}{-0} (\textcolor{fpcolor}{+10})   & \textcolor{fncolor}{-0} (\textcolor{fpcolor}{+17})   & \textcolor{fncolor}{-0} (\textcolor{fpcolor}{+28})   & \textcolor{fncolor}{-0} (\textcolor{fpcolor}{+9})   & \textcolor{fncolor}{-0} (\textcolor{fpcolor}{+4})   & \textcolor{fncolor}{-0} (\textcolor{fpcolor}{+2})    & 0.110 & \textbf{1.000} & 0.180   \\
    Qwen2.5-3B-FT     & \textcolor{fncolor}{-0} (\textcolor{fpcolor}{+17})  & \textcolor{fncolor}{-0} (\textcolor{fpcolor}{+38})   & \textcolor{fncolor}{-0} (\textcolor{fpcolor}{+9})   & \textcolor{fncolor}{-0} (\textcolor{fpcolor}{+18})   & \textcolor{fncolor}{-0} (\textcolor{fpcolor}{+28})  & \textcolor{fncolor}{-0} (\textcolor{fpcolor}{+9})   & \textcolor{fncolor}{-0} (\textcolor{fpcolor}{+4})  & \textcolor{fncolor}{-0} (\textcolor{fpcolor}{+3})    & 0.074 & \textbf{1.000}  & 0.138     \\
    DeepSeek-Llama3     & \textcolor{fncolor}{-0} (\textcolor{fpcolor}{+16})  & \textcolor{fncolor}{-1} (\textcolor{fpcolor}{+36})   & \textcolor{fncolor}{-0} (\textcolor{fpcolor}{+8})   & \textcolor{fncolor}{-0} (\textcolor{fpcolor}{+17})   & \textcolor{fncolor}{-0} (\textcolor{fpcolor}{+28})  & \textcolor{fncolor}{-0} (\textcolor{fpcolor}{+8})   & \textcolor{fncolor}{-0} (\textcolor{fpcolor}{+4})  & \textcolor{fncolor}{-0} (\textcolor{fpcolor}{+3})    & 0.133 & 0.941  & 0.233     \\
    DeepSeek-Qwen2.5     & \textcolor{fncolor}{-0} (\textcolor{fpcolor}{+14})  & \textcolor{fncolor}{-1} (\textcolor{fpcolor}{+36})   & \textcolor{fncolor}{-0} (\textcolor{fpcolor}{+8})   & \textcolor{fncolor}{-0} (\textcolor{fpcolor}{+16})   & \textcolor{fncolor}{-0} (\textcolor{fpcolor}{+28})  & \textcolor{fncolor}{-0} (\textcolor{fpcolor}{+9})   & \textcolor{fncolor}{-0} (\textcolor{fpcolor}{+4})  & \textcolor{fncolor}{-0} (\textcolor{fpcolor}{+3})    & 0.153 & 0.947  & 0.263     \\
    \midrule
    \SysName(Llama-8B)    & \textcolor{fncolor}{-7} (\textcolor{fpcolor}{+0})  & \textcolor{fncolor}{-5} (\textcolor{fpcolor}{+5})   & \textcolor{fncolor}{-1} (\textcolor{fpcolor}{+0})   & \textcolor{fncolor}{-3} (\textcolor{fpcolor}{+3})   & \textcolor{fncolor}{-16} (\textcolor{fpcolor}{+1})   & \textcolor{fncolor}{-7} (\textcolor{fpcolor}{+1})   & \textcolor{fncolor}{-3} (\textcolor{fpcolor}{+0})   & \textcolor{fncolor}{-2} (\textcolor{fpcolor}{+0})     & 0.891 & 0.651  &  0.752   \\
    \SysName(Qwen2.5-7B)    & \textcolor{fncolor}{-6} (\textcolor{fpcolor}{+0})  & \textcolor{fncolor}{-5} (\textcolor{fpcolor}{+4})   & \textcolor{fncolor}{-1} (\textcolor{fpcolor}{+0})   & \textcolor{fncolor}{-2} (\textcolor{fpcolor}{+3})   & \textcolor{fncolor}{-16} (\textcolor{fpcolor}{+1})   & \textcolor{fncolor}{-7} (\textcolor{fpcolor}{+0})   & \textcolor{fncolor}{-3} (\textcolor{fpcolor}{+0})   & \textcolor{fncolor}{-2} (\textcolor{fpcolor}{+0})     & 0.915 & 0.672  &  0.775   \\
    \SysName(Qwen2.5-3B-FT)    & \textcolor{fncolor}{-6} (\textcolor{fpcolor}{+0})  & \textcolor{fncolor}{-5} (\textcolor{fpcolor}{+2})   & \textcolor{fncolor}{-1} (\textcolor{fpcolor}{+0})   & \textcolor{fncolor}{-2} (\textcolor{fpcolor}{+3})   & \textcolor{fncolor}{-16} (\textcolor{fpcolor}{+1})   & \textcolor{fncolor}{-7} (\textcolor{fpcolor}{+0})   & \textcolor{fncolor}{-2} (\textcolor{fpcolor}{+0})  & \textcolor{fncolor}{-2} (\textcolor{fpcolor}{+0})   & 0.937 & 0.685  &  0.791   \\
    \SysName(DeepSeek-Llama3)    & \textcolor{fncolor}{-5} (\textcolor{fpcolor}{+0})  & \textcolor{fncolor}{-4} (\textcolor{fpcolor}{+1})   & \textcolor{fncolor}{-1} (\textcolor{fpcolor}{+0})   & \textcolor{fncolor}{-2} (\textcolor{fpcolor}{+2})   & \textcolor{fncolor}{-16} (\textcolor{fpcolor}{+0})   & \textcolor{fncolor}{-6} (\textcolor{fpcolor}{+0})   & \textcolor{fncolor}{-2} (\textcolor{fpcolor}{+0})  & \textcolor{fncolor}{-2} (\textcolor{fpcolor}{+0})   & 0.969 & 0.714  &  0.822   \\
    \SysName(DeepSeek-Qwen2.5)    & \textcolor{fncolor}{-4} (\textcolor{fpcolor}{+0})  & \textcolor{fncolor}{-4} (\textcolor{fpcolor}{+1})   & \textcolor{fncolor}{-1} (\textcolor{fpcolor}{+0})   & \textcolor{fncolor}{-2} (\textcolor{fpcolor}{+2})   & \textcolor{fncolor}{-16} (\textcolor{fpcolor}{+0})   & \textcolor{fncolor}{-5} (\textcolor{fpcolor}{+0})   & \textcolor{fncolor}{-2} (\textcolor{fpcolor}{+0})  & \textcolor{fncolor}{-1} (\textcolor{fpcolor}{+0})   & \textbf{0.970} & 0.737  &  \textbf{0.838}   \\
    \midrule
         \multicolumn{12}{c}{Evidence Selection Accuray} \\ \midrule 
    \# of Avg. Evidence (GT) & 2.59 & 2.30 & 3.10 & 3.37 & 1.79 & 1.00  & 1.00 & 3.00 & 1.000 & 1.000 & 1.000\\ \midrule 
    \SysName(Llama-8B) & \textcolor{fncolor}{-0.53} (\textcolor{fpcolor}{+0}) & \textcolor{fncolor}{-0.38} (\textcolor{fpcolor}{+0.10}) & \textcolor{fncolor}{-0.10} (\textcolor{fpcolor}{+0}) & \textcolor{fncolor}{-0.32} (\textcolor{fpcolor}{+0}) & \textcolor{fncolor}{-0.82} (\textcolor{fpcolor}{+0.07}) & \textcolor{fncolor}{-0.56} (\textcolor{fpcolor}{+0.11}) & \textcolor{fncolor}{-0.50} (\textcolor{fpcolor}{+0}) & \textcolor{fncolor}{-0.22} (\textcolor{fpcolor}{+0}) & 0.981 & 0.808 & 0.886 \\
    \SysName(Qwen2.5-7B) & \textcolor{fncolor}{-0.41} (\textcolor{fpcolor}{+0}) & \textcolor{fncolor}{-0.38} (\textcolor{fpcolor}{+0.05}) & \textcolor{fncolor}{-0.10} (\textcolor{fpcolor}{+0}) & \textcolor{fncolor}{-0.32} (\textcolor{fpcolor}{+0}) & \textcolor{fncolor}{-0.82} (\textcolor{fpcolor}{+0.07}) & \textcolor{fncolor}{-0.56} (\textcolor{fpcolor}{+0}) & \textcolor{fncolor}{-0.50} (\textcolor{fpcolor}{+0}) & \textcolor{fncolor}{-0.22} (\textcolor{fpcolor}{+0}) & 0.992 & 0.816 & 0.896 \\
    \SysName(Qwen2.5-3B-FT) & \textcolor{fncolor}{-0.41} (\textcolor{fpcolor}{+0}) & \textcolor{fncolor}{-0.23} (\textcolor{fpcolor}{+0.03}) & \textcolor{fncolor}{-0.10} (\textcolor{fpcolor}{+0}) & \textcolor{fncolor}{-0.32} (\textcolor{fpcolor}{+0}) & \textcolor{fncolor}{-0.82} (\textcolor{fpcolor}{+0}) & \textcolor{fncolor}{-0.56} (\textcolor{fpcolor}{+0}) & \textcolor{fncolor}{-0} (\textcolor{fpcolor}{+0}) & \textcolor{fncolor}{-0.22} (\textcolor{fpcolor}{+0}) & 0.998 & 0.853 & 0.920 \\
    \SysName(DeepSeek-Llama3) & \textcolor{fncolor}{-0.34} (\textcolor{fpcolor}{+0}) & \textcolor{fncolor}{-0.18} (\textcolor{fpcolor}{+0.02}) & \textcolor{fncolor}{-0.10} (\textcolor{fpcolor}{+0}) & \textcolor{fncolor}{-0.32} (\textcolor{fpcolor}{+0}) & \textcolor{fncolor}{-0.82} (\textcolor{fpcolor}{+0}) & \textcolor{fncolor}{-0.56} (\textcolor{fpcolor}{+0.11}) & \textcolor{fncolor}{-0.25} (\textcolor{fpcolor}{+0}) & \textcolor{fncolor}{-0.22} (\textcolor{fpcolor}{+0}) & 0.992 & 0.845 & 0.913 \\
    \SysName(DeepSeek-Qwen2.5) & \textcolor{fncolor}{-0.27} (\textcolor{fpcolor}{+0}) & \textcolor{fncolor}{-0.18} (\textcolor{fpcolor}{+0.02}) & \textcolor{fncolor}{-0.10} (\textcolor{fpcolor}{+0}) & \textcolor{fncolor}{-0.32} (\textcolor{fpcolor}{+0}) & \textcolor{fncolor}{-0.82} (\textcolor{fpcolor}{+0}) & \textcolor{fncolor}{-0.44} (\textcolor{fpcolor}{+0}) & \textcolor{fncolor}{-0} (\textcolor{fpcolor}{+0}) & \textcolor{fncolor}{-0.11} (\textcolor{fpcolor}{+0}) & \textbf{0.999} & \textbf{0.876} & \textbf{0.934} \\
    \bottomrule
    \end{tabular}
\end{table*}

\subsection{Accuracy }
\label{eval:accuracy}

To quantitatively evaluate \SysName’s accuracy, we randomly sampled a subset of 136 applications, representing approximately 5\% of the real-world dataset.
We then engaged security analysts to manually review these samples.
Specifically, the analysts annotated each application with ground-truth labels corresponding to its specific risk categories, along with the associated evidentiary snippets.
\subsubsection{Accuracy of Risk Identification}
\SysName on \textit{DeepSeek-Distilled} model series demonstrated superior detection efficacy compared to \textit{Llama3} and \textit{Qwen2.5} models. 
Within the DeepSeek model series, the \textit{DeepSeek-Distilled-Qwen2.5-7B} exhibits greater task-specific efficacy. \SysName on \textit{DeepSeek-Distilled-Qwen2.5-7B} exhibited further improvements, attaining 72.1\% accuracy, 97.0\% precision, 73.7\% recall, and an F1-score of 83.8\%. The fundamental reason lies in the visualizable thinking processes of the \textit{DeepSeek} models, which ensure more effective optimization and refinement during prompt engineering, thereby enabling the design of highly targeted prompt instructions to mitigate model hallucinations. Additionally, since the operational context is situated in Chinese environments, the \textit{DeepSeek-Distilled-Qwen2.5} model, trained specifically on Chinese corpora, demonstrates superior task performance compared to counterparts lacking equivalent linguistic specialisation.

Analysis of false negatives and positives revealed two main causes for errors. First, missing or incomplete metadata made issue detection challenging. Second, risk categories like Content Risk and Illegal Features rely on user feedback, often collected after extensive use, complicating data capture. User feedback accuracy is often compromised by respondents' lack of expertise and emotionally charged responses, which introduce noise into sentiment interpretation and risk prioritization. For example, applications providing free calls or SMS services received positive feedback despite facilitating Illegal Features, sometimes skewing \SysName's assessment. 

To further assess its real-world performance, we deployed \SysName on a dataset of 2,096 previously unlabeled applications.
\SysName automatically flagged 106 applications as potentially risky.
Subsequent manual verification by security experts confirmed that 87 of these flagged instances represented verifiable risks, yielding a precision of 82.1\% (87/106).
This demonstrates the system's strong capability for accurate risk identification in a practical deployment scenario.

\subsubsection{Accuracy of Evidence Retrieving}
To further validate the accuracy of evidence selection, we quantified the precision of the evidence chosen during the construction of evidence chains by the system. As shown in Table~\ref{tab:eva1}, 
\SysName using the \textit{DeepSeek-Distilled-Qwen2.5-7B} model recorded a precision of 99.9\%, a recall of 87.6\%, and an F1-score of 93.4\%. 
This demonstrates that \SysName can effectively retrieve features related to the risk from massive datasets, significantly reducing the cost of manual analysis. 

\begin{figure}[tbp]
    \centering
    \includegraphics[width=0.42\textwidth]{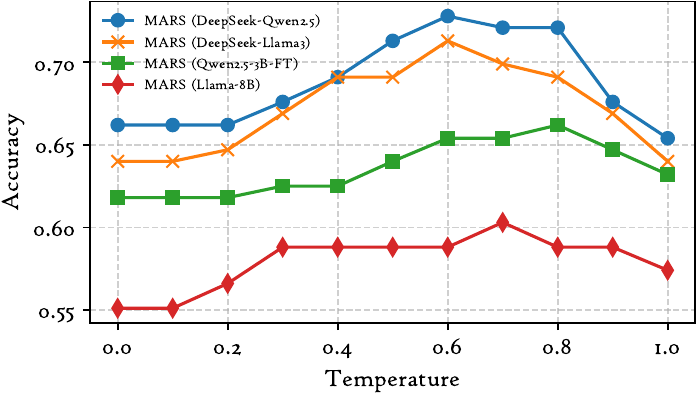}
    \vspace{-0.1in}
    \caption{Selecting Models' Temperature}\label{fig:model_temperature}
    \vspace{-0.1in}
\end{figure}

\subsubsection{Selecting Temperature} To investigate the impact of the temperature parameter on model performance in this context, experimental evaluations were conducted across a temperature range of 0.0 to 1.0 (in increments of 0.1) using a subset of 136 applications. For each temperature setting, five independent trials were conducted, and the average accuracy was computed to ensure robustness and mitigate variance. Results demonstrate task-specific optimal configurations: the DeepSeek model achieved peak performance at 0.6, Qwen exhibited optimal detection efficacy at 0.8, while the \textit{Llama} architecture demonstrated superior results at its default setting of 0.7, suggesting inherent architectural dependencies in temperature sensitivity. Detailed performance metrics across temperature configurations are provided in Figure~\ref{fig:model_temperature}.

\subsubsection{Ablation Study} 
To validate the impact of individual components on \SysName's detection accuracy, we conducted ablation studies using a baseline configuration devoid of both mechanisms (Indentification Tree module and Delisted application Validation module). Three experimental configurations were comparatively evaluated: 1) the system without identification tree, 2) the system excluding historical delisted application validation, and 3) the full system incorporating both mechanisms. All the above experiments were conducted under identical experimental conditions and on the same dataset, the only variable was the underlying mechanism used. 

\begin{table}[tbp]
\footnotesize
    \centering
    \caption{Break-down Analysis of Validation Mechanism}
	\begin{tabular}{c|ccc}
	    \toprule Mechanism  & \makecell{False\\Positives} & \makecell{Precision} & \makecell{F1-Score}\\
		\midrule 
        {\makecell{w/o Both Mechanism}} & 65 & 0.356 & 0.419 \\
        {\makecell{w/o Risk Identification Tree}} & 39 & 0.614 & 0.626 \\
	{\makecell{w/o Similar Risky Applications}} & 17 & 0.706 & 0.764\\
        {\makecell{w/ Both Mechanism}} & 3 & 0.970 & 0.838\\
		\bottomrule
    \end{tabular}
    \label{tab:breakdown}
\end{table}

Furthermore, as shown in Table~\ref{tab:breakdown}, further ablation studies prove that the verification methods described in \S\ref{subsec:validation} effectively reduce false positives and enhance the accuracy of the analysis. Experimental results demonstrate that implementing solely the Validation with Delisted Apps mechanism achieved a reduction of 26 false positive (FP) cases, while the identification Tree component alone reduced FP instances by 48. This performance discrepancy can be primarily attributed to the exclusion of erroneous risk classifications during the identification tree filtering process, which effectively mitigates model hallucination tendencies by eliminating logically inconsistent intermediate hypotheses prior to final risk adjudication.

\subsection{Computational Cost}\label{eval:efficiency}
The overhead introduced by the Large Language Model is a primary component of the system's operational cost. Our analysis quantifies this by measuring token consumption and processing time across the two phases where the LLM is utilized: risk identification and evidence chain generation.
As shown in Table~\ref{tab:eva1} and Table~\ref{tab:token_and_time_models}, following our evaluation in RQ1, the \textit{DeepSeek-Distilled-Qwen2.5-7B} model was selected for implementation in \SysName, as it provides the optimal balance between detection efficacy and resource expenditure. The experimental results, presented in Table \ref{tab:token_and_time}, show that the risk identification phase has an average token consumption of 500.5 and a latency of 13.87 seconds. The subsequent evidence generation phase naturally incurs a higher computational cost, as the synthesis and integration of diverse textual evidence demand a larger context window and generate more extensive output.

\begin{table}[tbp]
\footnotesize
    \centering
    \caption{Token and Time Consumption of Different LLMs} \label{tab:token_and_time_models}
    \begin{tabular}{l|cc}
        \toprule
        Model & \# of Tokens  & Time (s)  \\ 
        \midrule
        \SysName(Llama-8B) & 1623.2  &  12.13    \\
        \SysName(Qwen2.5-7B) & 1409.4  &  10.28    \\
        \SysName(Qwen2.5-3B-Fine-tuned &  1516.6 &  10.47   \\
        \SysName(DeepSeek-Distilled-Llama-8B) &  3387.1 &  26.87  \\
        \SysName(DeepSeek-Distilled-Qwen-7B) &  3264.3 &  24.68   \\
        \bottomrule
        \end{tabular}
\end{table}

\begin{table}[tbp]
\footnotesize
    \centering
    \caption{Overall Token and Time Consumption} \label{tab:token_and_time}
        \begin{tabular}{c|cc|cc|cc}
        \toprule
          & \multicolumn{2}{c|}{Risk Identification} & \multicolumn{2}{c|}{Evidence Generation} & \multicolumn{2}{c}{Summary}\\
          & Tokens \# & Time (s) & Token \# & Time (s) & Tokens \# & Time (s)\\ \midrule
        R1 &  572.1  & 21.20 &  3549.1 &  22.67  & 4121.2 & 43.87\\ 
        R2 &  662.1  & 18.21 & 2654.7  &  12.17  & 3316.8 & 30.38\\ 
        R3 &  265.7  & 8.51 &  2488.6 &  9.57  & 2754.3 & 18.08  \\ 
        R4 &  469.8  & 10.42 &  3601.2 &  19.20  & 4071.0 & 29.62\\ 
        R5 &  209.2  & 6.11 &  2202.5 &  6.10  & 2411.7 &  12.21\\ 
        R6 &  595.6  & 13.22 &  1901.2 &  7.20  & 2496.8 & 20.42 \\ 
        R7 &  799.2  & 25.10 &  3702.1 &  6.22  & 4501.3 &  31.32\\ 
        R8 &  430.5  & 8.22 &  2006.2 &  3.32  & 2430.7 & 11.54 \\ 
        \midrule
        Avg. & 500.5   & 13.87   &  2763.8 & 10.81  & 3264.3 & 24.68 \\ 
       \bottomrule
        \end{tabular}
    \vspace{-0.1in}
\end{table}

Aggregating both stages, the end-to-end analysis of a single application requires an average of 3,264.3 tokens and 24.68 seconds. It is noteworthy that the DeepSeek model's token consumption is higher than that of other tested models (e.g., Qwen, Llama); this is an expected characteristic resulting from its explicit inclusion of detailed reasoning traces in the output, which enhances interpretability. Despite this, the model's low per-token cost makes it highly cost-effective. The total expenditure for analyzing all 2,232 applications was approximately \$33.50, yielding an average cost of \$0.015 per app. This demonstrates that the overhead of \SysName, in both time and cost, is well within acceptable limits for practical deployment scenarios.

\subsection{Applicability}\label{eval:applicability}

\subsubsection{Report Interpretability}

The interpretability of reports is a critical feature of \SysName. We demonstrate this through a case study that assesses how the report's structure enhances both the efficiency of risk identification and the clarity of the information presented to an analyst.
\SysName's reports are architected for human comprehension, consisting of three synergistic components, as illustrated in the ``I Am Music Library'' case study (Table~\ref{tab:casestudy}):
1) Metadata: Establishes a baseline identity for the application (name, developer, etc.).
2) Evidence Snippets and Risk Descriptions: Forms the core of the report, providing direct, intelligible evidence and a summary of the potential violation.
3) Similar Delisted Apps: Augments the analysis by retrieving and displaying similar applications that were previously delisted, offering valuable contextual precedent.

\SysName streamlines the traditional app profiling workflow by automating the burdensome tasks of information gathering and threshold configuration. Instead of relying on predefined rules, our system directly extracts and synthesizes evidence from multiple data sources, interpreting it with embedded expert knowledge to provide rich context. The final output is a highly interpretable report—combining direct evidence, concise descriptions, and relevant historical precedents—that enables engineers to rapidly validate potential risks and focus their efforts on verification and mitigation.

\subsubsection{User Study}
We conducted a user study with 20 experienced security practitioners from a partner company to evaluate \SysName's real-world impact. Participants reviewed five sample risk reports generated by our system and completed a survey, as Table~\ref{tab:questionnarie} shows, assessing its efficiency and the quality of its output.
A key outcome of the study was identifying the primary bottlenecks that impede manual review. Participants consistently highlighted three major challenges: the laborious process of gathering information from multiple sources, the frequent need for manual intervention to verify vague alerts, and the difficulty of reproducing issues due to a lack of context.

\begin{figure}[!t]
    \centering

    \begin{subfigure}[b]{0.49\linewidth}
        \centering
        \includegraphics[width=\linewidth]{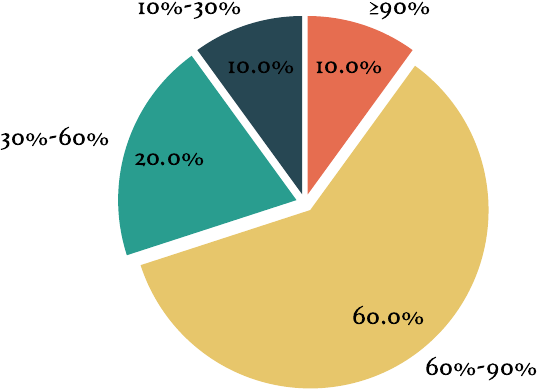}
        \caption{Efficiency}
        \label{fig:pie_chart}
    \end{subfigure}
    \begin{subfigure}[b]{0.49\linewidth}
        \centering
        \includegraphics[width=\linewidth]{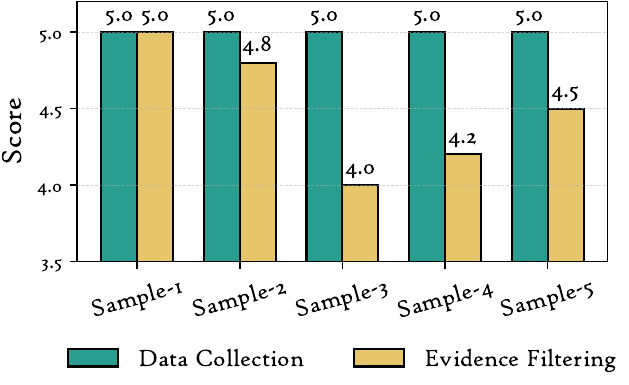}
        \caption{Accuracy}
        \label{fig:grouped_chart}
    \end{subfigure}

    \vspace{0.6em}

    \begin{subfigure}[b]{\linewidth}
        \centering
        \includegraphics[width=0.92\linewidth]{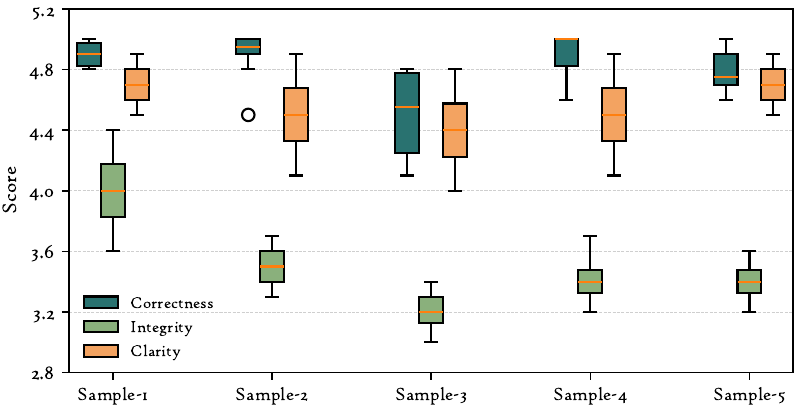}
        \caption{Applicability}
        \label{fig:box_chart}
    \end{subfigure}
    \vspace{-0.1in}

    \caption{Summary of User Study}
    \label{fig:survey_results}
    \vspace{-0.1in}
\end{figure}

\SysName is designed to directly address these bottlenecks. Its automated data integration and filtering pipeline replaces manual information gathering. The system's hybrid reasoning reduces false positives, minimizing the need for manual intervention. While it does not fully automate issue reproduction, its structured reports provide actionable forensic traces that help engineers rapidly localize root causes.
The practical impact of these improvements was validated by the study participants. When asked to estimate the efficiency gains, over 50\% of participants projected that \SysName would reduce their review time by 60\% to 90\% (Figure~\ref{fig:pie_chart}). This highlights a substantial improvement in operational efficiency over conventional practices.

\begin{table}[tbp]
\footnotesize
    \centering
    \caption{User Study Questionnarie (Partial)} \label{tab:questionnarie}
    \begin{tabular}{p{5cm}|p{3cm}}
        \toprule
        \multicolumn{1}{c|}{Questions} & \multicolumn{1}{c}{Options}   \\ 
        \midrule
        What's your job position? &  Developer/ Auditor/ Others   \\
        How many years of professional experience do you have? &  Less than 1 / 1-3 / 4-6 / 7-10 / 10+\\
        How much time you need to review an app? & 1-5 / 6-10 / 11-15 / 15+ (min) \\
        What is the primary bottleneck in your current review workflow? & Insufficient tool support /\newline Complicated data processing / \newline Lack of domain knowledge / \newline Other (please specify) \\
        Does \SysName help reduce your app review time? & Yes/No \\
        Approximately what percentage of time was saved by using the system? & 10-30\%/ 30-60\%/ 60-90\%/ 90\%+ \\
        Please rate the reports generated by \SysName on ``Accuracy,'' ``Correctness,'' ``Integrity,'' and ``Clarity.'' & Very Low / Low / Medium /\newline High / Very high \\
        \bottomrule
        \end{tabular}
\end{table}

The quantitative evaluation required participants to score the quality of risk reports from \SysName. We first assessed the perceived quality of the underlying evidence by asking participants to rate its Completeness (reflecting data collection) and Relevance (reflecting evidence filtering). As shown in Figure~\ref{fig:grouped_chart}, the system received near-perfect average scores (5.0 and 4.5, respectively), indicating that engineers found the reports to be both comprehensive and well-focused on key risk indicators.

Next, we evaluated the final evidence chain itself across three dimensions: Correctness (4.8/5), Clarity (4.6/5), and Integrity (3.5/5), with detailed results in Figure~\ref{fig:box_chart}. While the scores for correctness and clarity were exceptionally high, the lower score for integrity warrants discussion. This result was largely due to a single challenging case, Sample-3 ("Illegal Features"), where the app's latent malicious behavior impeded full data acquisition. This known limitation resulted in a less complete evidence chain for that specific case, which the participants accurately identified.
Overall, these findings demonstrate that \SysName is highly user-friendly. By generating reports that engineers perceive as complete, relevant, correct, and clear, the system provides actionable intelligence that allows security teams to contextualize and manage risks more efficiently.

\section{Discussion and Conclusion}

\subsection{Discussion}
Our work also presents several clear avenues for future research and development:
\begin{list}{\labelitemi}{\leftmargin=1.6em}
 \setlength{\topmargin}{0pt}
 \setlength{\itemsep}{0em}
 \setlength{\parskip}{0pt}
 \setlength{\parsep}{0pt}
    \item \textbf{Automated Behavior Reproduction.} A key limitation highlighted by our user study is the continued reliance on manual effort for risk confirmation.
    While \SysName provides interpretable guidance, a significant next step is to extend the system to generate explicit, actionable instructions for reproducing identified behaviors, bridging the gap between evidence interpretation and final validation.

    \item \textbf{Expanding Risk and Feature Coverage.} The system's knowledge tree is inherently modular. Future work will leverage this design to expand support for a broader range of application features. We also plan to enhance the automated mechanism that constructs new trees, allowing the system to adapt to emerging risk categories and the evolving threat landscape.

    \item \textbf{Optimizing LLM Performance and Efficiency.} Our focus on smaller, computationally efficient LLMs introduces trade-offs, including challenges with structured output generation and susceptibility to hallucination. We plan to explore advanced prompt engineering and tuning techniques to mitigate these issues and to investigate the optimal balance between model size, performance, and operational cost.
\end{list}

\subsection{Conclusion}

This paper introduced \SysName, an end-to-end framework that leverages Large Language Models for automated mobile app risk profiling. By combining the reasoning power of LLMs with a knowledge-guided approach, \SysName effectively identifies risks from multi-source data and generates interpretable evidence chains. Our real-world evaluation demonstrated its high accuracy (F1-score of 0.838) and efficiency, with a user study confirming a 60\%–90\% time savings in typical review workflows.

\begin{acks}
To Robert, for the bagels and explaining CMYK and color spaces.
\end{acks}

\bibliographystyle{ACM-Reference-Format}
\bibliography{Main}


\end{document}